\documentclass[12pt]{article}
\usepackage{a4}
\usepackage[dvips]{graphicx}
\usepackage{latexsym}
\usepackage{euscript}
\newcommand{\ifig}[1]{\includegraphics[height=8cm,width=14cm]{#1}}
\newcommand{\bc}{\begin{center}}
\newcommand{\ec}{\end{center}}
\newcommand{\be}{\begin{equation}}
\newcommand{\ee}{\end{equation}}
\newcommand{\bea}{\begin{eqnarray}}
\newcommand{\eea}{\end{eqnarray}}
\newcommand{\ba}{\begin{eqnarray}}
\newcommand{\ea}{\end{eqnarray}}

\newcommand{\AC}{\cal{A}}
\newcommand{\VC}{\cal{V}}

\newcommand{\simge}{\ \lower-1.2pt\vbox{\hbox{\rlap{$>$}\lower5pt
\vbox{\hbox{$\sim$}}}}\ }

\begin{document}
\centerline{\large {\bf{Remarks on Lattice Gauge Fixing}}}
\vskip 1.0cm
\centerline{S. Petrarca}
\centerline{\small Dipartimento di Fisica, Universit\`a di Roma "La
Sapienza",}
\centerline{\small P.le A. Moro 2, I-00185 Roma, Italy.}
\centerline{\small INFN, Sezione di Roma 1,
P.le A. Moro 2, I-00185 Roma, Italy.}
\centerline{\small P.le A. Moro 2, I-00185 Roma, Italy.}
\vskip 15mm

 In this talk I briefly comment on the conventional lattice gauge fixing adopting 
 a critical, even though
 constructive, numerical point of view. 

\section{Standard Landau}
On the lattice the usual procedure accepted to compute 
gauge dependent matrix elements is summarized in 
the following formula defining
the expectation value of a gauge dependent operator 
${\cal O}$:
\be\label{eq:omedio}
\langle{\cal O}\rangle=\frac{1}{Z}
\int dU {\cal O}({U^{G}})
e^{-\beta S(U)}\; ,
\ee
 where $S(U)$ is the Wilson lattice gauge invariant action, 
$G$ is the gauge transformation projecting the links
in the Landau gauge
\be
\partial_\mu A^G_\mu=0 \; + \;{\rm periodic}\; {\rm boundary}\; {\rm condition} 
\label{eq:landau}
\ee
with the gauge rotation given by ${U^{G}_\mu (x)}=G(x) U_\mu (x) G^{\dagger} (x+\mu)$
 and the gluon field  defined on the lattice in the standard way:
\be
A_{\mu} (x) 
\equiv  \left[{{U_{\mu} (x) - U_{\mu}^{\dagger} (x)}\over
{2 i a g_0}}\right]_{Traceless}\; .
\label{eq:standard}
\ee 

In the lattice gauge theories where the links belong to a compact group,
the gauge fixing is necessary only in the case of the measure of a gauge 
dependent operator. An expression similar to  (\ref{eq:omedio}) but without
having gauge fixed the links, defines the expectation value
of a gauge independent operator.    
Moreover, in a lattice simulation there is no need to compute the 
Faddeev-Popov determinant because
 the correct adjustment of the measure, necessary in the case of gauge fixing,
 is obtained rotating the links
 in the chosen gauge.
Therefore the complexity of the ghost technique is replaced by 
the numerical evaluation of the gauge transformations. The price to pay is
the large amount of computer time spent to obtain numerically the gauge
transformations. From  a numerical point of a view the values of
a gauge dependent operators strongly fluctuate around zero if the gauge 
has  not been
fixed. In the case of an imperfect or inadequate  gauge fixing the measure
of a gauge dependent operator is  affected by additional fluctuations 
to be summed up  to the intrinsic statistical noise.

The necessary steps bringing to the
computation of the integral  (\ref{eq:omedio}) can be described as 
follows:
\begin{itemize}
\item  A set of $N$ thermalized configurations 
$\{U\}$ is generated with periodic boundary conditions according to the gauge invariant 
weight $e^{-S_W(U)}$;
\item  For each  $\{U\}$ a numerical algorithm compute the gauge transformation 
$G$;
\item The expectation value of an operator is
 given by the mean value of the values taken by  the operator
 on the gauge rotated configurations: 
\end{itemize}
\be
\langle{\cal O}\rangle^{Latt}=\frac{1}{N}\sum_{\{conf\}}{\cal O}({U}^G)\;\; .
\ee

 The gauge fixing algorithm is based on the minimization of a functional
 $F_U[G]$ constructed in such a way that its extrema are the gauge fixing
 transformations corresponding to the gauge condition.
 The $F$ standard form  for the Landau gauge is:
 \begin{equation}
 F_U[G]= -Re \; Tr \; \sum_{\mu,x} {U_{\mu}}^{G(x)} (x)
 \label{eq:effel}
 \end{equation}
 and the transformations $G$ for which
 $\frac {\delta F} {\delta G} =0$ rotate the links in the gauge
 $\partial_\mu A^G_\mu=0 $.
 The algorithm sweeps all the lattice  many times
 and it stops when a prefixed quality factor is reached.
 
 It is remarkable that the eq. (\ref{eq:effel}) 
 does not correspond to the natural discretization of the continuum
 functional
 \begin{equation}
F_A [G] \equiv -\ Tr \ \int d^4 x \ \left( A_{\mu}^G (x) A_{\mu}^G (x) \right) 
\equiv -\left( A^G, A^G \right) \equiv -||A^G||^2 \;,
\label{eq:effe}
\end{equation}
according to the lattice definition of the gluon field (\ref{eq:standard}) but
it differs from that by $O(a)$ terms.
The form in eq. (\ref{eq:effel}) is adopted not only for its simplicity but
also because its minimization enforces
the following discretized version of the gauge condition
\begin{equation}
\Delta^G(x) \equiv \sum_{\mu=1}^{4} \ ( A^G_{\mu} (x) - A^G_{\mu} (x - \hat{\mu} 
))= 0  
\label{eq:landlat}
\end{equation}
where  $A_\mu$ must be related to the links by the standard
definition (\ref{eq:standard}).

In order to study the approach to the minimum, two 
quantities are usually monitored. The first one is $F[U^G]$ itself, which 
decreases monotonically and eventually reaches a plateau. The other one, 
denoted 
by $\theta$, is defined as follows:
\begin{equation}
\theta^G \equiv {1 \over V} \ 
\sum_{n} \theta^G(x) \equiv {1 \over V}
\ \sum_{n} Tr \ [ \Delta^G (x) (\Delta^G)^{\dagger} (x)],
 \simeq  \int d^4 x \  Tr(\partial_\mu A_{\mu}^G )^2 ,
\label{eq:thetalat}
\end{equation}
where $V$ is the lattice volume.

The function $\theta$  decreases (not 
strictly monotonically) approaching zero when $F_U[G]$ reaches its minimum.
The desired gauge fixing quality is determined stopping the computer code
when $\theta^G$ has achieved a preassigned value close to zero.
The choice of the gauge fixing quality is a delicate point in the case
of a simulation with a large volume and a high number of thermalized 
configurations. Of course, the better is the gauge fixing quality, the more
computer time is needed. Moreover it is impossible to know before computing
the gauge dependent correlation functions if the choice done is suitable.
So that, the stopping $\theta$ value is normally fixed 
on the basis of a practical  compromise between the estimated computer time
and the gauge fixing quality. Sometimes, in the case of calculations 
performed on computers with
 single precision floating point, the maximum gauge fixing quality is
limited by a value of the order of the floating point zero:
 $\theta \simeq  10^{-7}$, this value is usually enough to guarantee
 the stability of  gauge dependent correlators.


\section{ Lattice Gribov Copies}
  On the lattice  a conceptual and numerical difficulty 
  connected with  gauge fixing is 
   the existence of many different minima of the functional
 $F[U]$. The different gauge transformations
 determined by different minima are not equivalent each other
 and can be labelled with the value of the functional $F$. Of course
 it is unthinkable to succeed in reaching numerically 
 the absolute minimum. The search of the  $F$ minima is at least
 as difficult as to find the lowest state of energy of a spin glass
 system with  hamiltonian $F$.
 So that the condition (\ref{eq:landau}) does not fix the gauge in
 a complete way generating on the lattice  
 a problem analogous  to the Gribov copies in the 
 continuum \cite{Gribov,vanbaal}. However the analogy is only formal because it is not
 possible to establish a connection among continuum and lattice copies.
 Moreover it is also likely that (many) lattice Gribov copies
 are spurious solutions due to the discretization \cite{giusti}.
 
 Actually, it must be noted that the presence of Gribov copies in the
 lattice Landau gauge fixing is a very common phenomenon.
 For example, it has been shown that when the 
  minimization algorithm 
  includes the over-relaxation technique,
 varying the value of the 
 over-relaxation parameter $\omega$ 
 different lattice Gribov copies \cite{over} are generated.
 Of course there is no correlation between the convergence rate 
 and the value of $F$ associated with the particular Gribov copy found.
 
 The numerical effects of lattice Gribov copies can be divided 
 into two categories: the distortion of a measurement and the
 lattice Gribov noise.
 The typical example of a distortion due to the existence of
 Gribov copies is the measure of the photon propagator in compact
 $U(1)$ in the so called Coulomb phase. In this case the measure
 of the photon propagator as function of the momentum, performed using
 the gauge fixing in the
 standard way, was affected by a not regular behaviour \cite{plewnia}.
 This problem was associated with the distortional effects due to 
 the Gribov copies. In fact,
  after having chosen the gauge fixed configurations nearest
 to the minimum of the gauge functional, the photon propagator became
  a smooth
 momentum function. More recent studies  \cite{mitrj} show
 the details  of the Gribov copies dynamics and provide a practical
 procedure to eliminate their effects. 
 It is interesting  to note that in the case of the measure
 of the gluon propagator in $SU(3)$ there is no  signal in the literature about
 a similar problem (for a recent review see ref. \cite{mandula}).
  The numerical simulations are performed in the Landau gauge and
 the various authors claim that the effects of Gribov copies do not
 affect the measure.
 
 Anyway, in the
 normal case in which there is no distortion
 due to Gribov copies, there  should  be  an increase of the numerical
 fluctuations due to the incomplete gauge fixing associated with the copies.
 An attempt to study the properties of this noise has been done
 in ref. \cite{za} taking as an example the measurement of
 the lattice axial current $Z_A$. This quantity
 is particularly well suited to the study
 of the Gribov fluctuations, because
  it is a gauge independent quantity but it can be obtained
from chiral Ward identities in two distinct ways:
a gauge independent one, which consists in taking the matrix elements
between hadronic states without fixing the gauge in the simulation,
 and a gauge dependent one, which consists
in taking the matrix elements between quark states in the Landau gauge.
In  the intermediate steps of the numerical computation, the
 second procedure takes into account gauge dependent matrix elements
 potentially subjected to the Gribov noise.
  Hence,
there is an explicitly gauge invariant estimate of $Z_A$ which is
free of Gribov noise and which can be directly compared to the gauge
dependent, Gribov affected, estimate.

The results of the analysis can be summarized in the following way:
\begin{itemize}
\item  there is a clear evidence of residual gauge freedom 
 associated with lattice Gribov copies;
 
\item the lattice Gribov noise is not separable from the statistical
uncertainty of the Monte Carlo method.
\end{itemize}

The global effect is not dramatic because the $Z_A$ value obtained
with the gauge dependent methods (1.08(5)) is close to the gauge independent
evaluation (1.06(6)) and the jacknife errors are comparable.

\section{Gluon Field Definition}
In order to impose the Landau gauge on the lattice it is necessary
to define the gluon field $A_\mu$ in terms of the links. It is clear
that the definition given in eq. (\ref{eq:standard}) is far from unique and
it cannot be preferred, from the first principles, to
any other definition with analogous properties.
Moreover, in the general field theoretical framework
 any pair of operators differing from
each other by irrelevant terms, i.e. formally equal up to terms of order
$a$, will tend, to the same continuum operator, up to a constant.
It is has been shown in ref. \cite{arte} that this feature is 
satisfied at the non-perturbative level in
lattice QCD. In fact,  different definitions
of the gluon field, at the lattice level, give rise to Green's functions
proportional to each other, thus guaranteeing the uniqueness of the renormalized
continuum gluon field.
The relation between the two $A_\mu$ definitions can be expressed up to
 $O(a^2)$ terms in this way \cite{testa}:
\be
{A^{'}}_{\mu}(x)=C(g_0) A_{\mu}(x). \label{eq:relazione}
\ee
Therefore for a Green's functions insertions the following
ratio is expected to be a constant
\be
{{\langle \dots A'_{\mu}(x) \dots \rangle} \over {\langle \dots A_{\mu}(x)
\dots \rangle}}= C(g_0) \; . \label{green}
\ee
This relation has been checked numerically on the lattice
by measuring 
a set of Green functions related to the gluon propagator 
for $SU(3)$ in the Landau gauge with periodic
boundary conditions.
\begin{figure}
\bc
\ifig{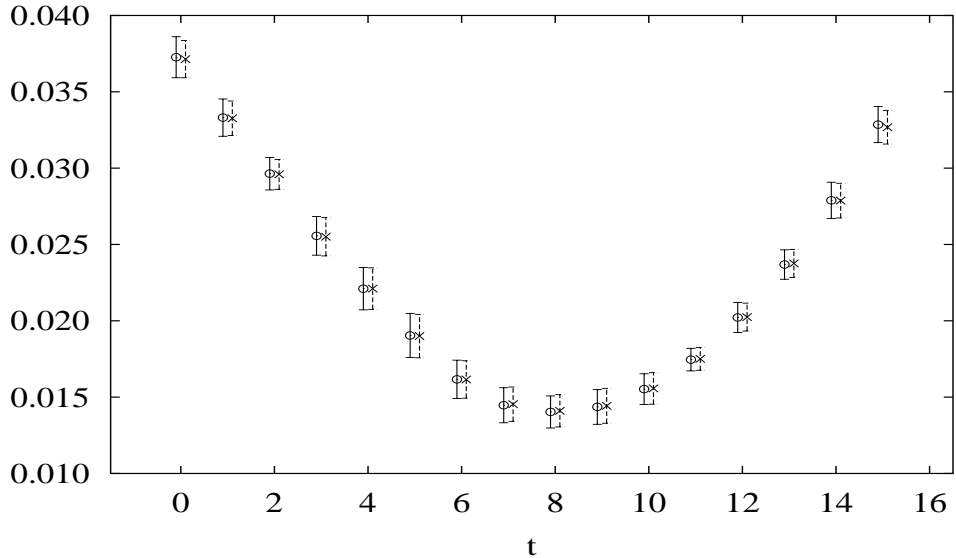}
\caption{\small{Comparison of the matrix elements of  $\langle{\AC}^{'}_i{\AC}^{'}_i\rangle(t)$ 
(crosses) and the rescaled $\langle{\AC}_i{\AC}_i\rangle~\cdot~C_i^2(g_0)$
(open circles) as function of time for 
a set of 50 thermalized $SU(3)$  configurations at $\beta=6.0$ with a 
volume $V\cdot T=8^3\cdot 16$. The data have been
slightly displaced in $t$ for clarity, the errors are jacknife.}}
\label{fig:amu}
\ec
\end{figure}
In Fig.~\ref{fig:amu} the Green functions $\langle {\AC}^{'}_i{\AC}^{'}_i\rangle$
and the rescaled one
$C_i^2(g_0) \langle {\AC}_i{\AC}_i\rangle$ are shown, where
\be
\langle {\AC}_i{\AC}_i\rangle (t) \equiv  \frac{1}{3 V^2}  
 \sum_{i}\sum_{{\bf x},{\bf y}} Tr \langle  A_i({\bf x},t)A_i({\bf y},0)\rangle 
\label{eq:AiAi}\\
\ee
and the operator $\langle {\AC}^{'}_i{\AC}^{'}_i\rangle (t)$
is obtained replacing in the same form the alternative definition:
\be
{A^{'}}_{\mu} (x) \
\equiv \ {{( (U_{\mu} (x))^2 - (U_{\mu}^{\dagger} (x))^2 )_{traceless}}
\over {4 i a g_0}}, \ \  \mu = 1, \ldots 4. \label{eq:seconda}
\ee
The remarkable agreement between these two quantities
confirms the proportionality shown in eq. (\ref{eq:relazione}).

\begin{figure}
\bc
\ifig{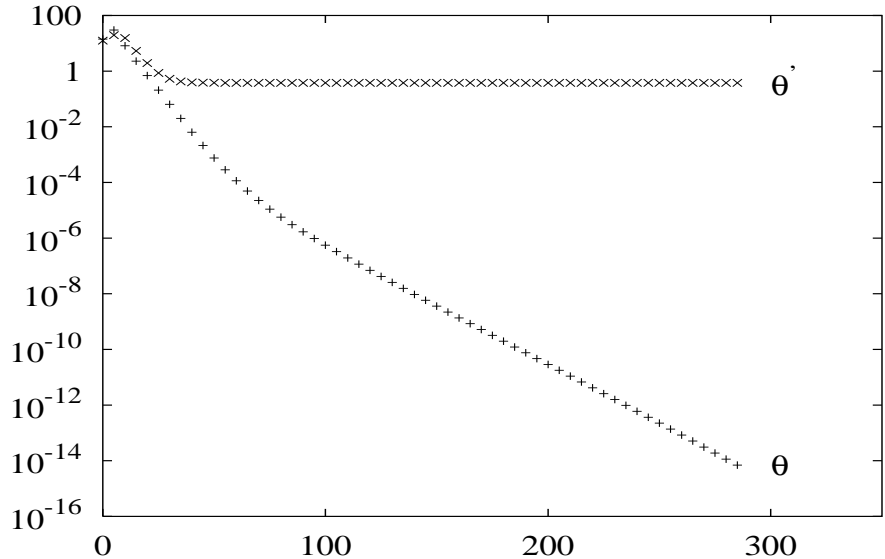}
\caption{\small{Typical behaviour of $\theta$ and $\theta^{'}$ vs gauge fixing sweeps at $\beta=6.0$
for a thermalized $SU(3)$ configuration $8^3\cdot 16$.}}
\label{fig:p60t}
\ec
\end{figure}

From the numerical point of view, however, the various definitions are
not interchangeable. In fact let me suppose to fix 
the gauge
of a thermalized configuration stopping the gauge-fixing sweeps
when $\theta \leq 10^{-14}$ and then define $\theta^{'}$ as having the
same functional form of $\theta$, as in eq. (\ref{eq:thetalat}), but
with $A_\mu$ replaced by $A^{'}_\mu$ given in eq. (\ref{eq:seconda})
The values of $\theta$ and $\theta^{'}$ during the minimization
of $F$ are reported in Fig. \ref{fig:p60t}, for a typical thermalized
configuration, as functions of the lattice
sweeps of the numerical gauge-fixing algorithm.
 As clearly seen
$\theta^{'}$ does not follow the same decreasing behaviour as $\theta$:
after an initial decrease, $\theta^{'}$ goes to a constant value, many orders of
magnitude higher than the corresponding value of $\theta$.
This difference, already noted in ref. \cite{giusti},
 between the behavior of $\theta$ and
$\theta^{'}$ 
could cast some doubts on the lattice gauge-fixing
procedure and on the corresponding continuum limit of gauge dependent operators.

This paradoxical situation is due to the fact that $\theta^{'}$ is an operator
 and the lattice can attribute a value to it only after averaging it
 over the gauge fixed configurations of the thermalized set. 
 Hence the comparison reported in Fig. \ref{fig:p60t} is devoid of meaning
 because it is done comparing the values obtained by a single configuration.
Moreover the behavior shown in Fig. \ref{fig:p60t} can be readily understood
in the following way.
The operator $\theta $, defined in eq. (\ref{eq:thetalat})
 ($\theta^{'}$),  is computed in the lattice units taking the definition  
eq.(\ref{eq:standard})  (eq.(\ref{eq:seconda}))
 without the powers of $a$ to the denominator.
Then in the continuum variables
$\theta =\frac{a^4}{\VC}\int d^4x (\partial_\mu A_\mu(x))^2$
 where $\VC$ is the 4-volume in physical
units  (analogously  for  $\theta^{'}$).
Hence, while $\theta$ vanishes configuration by configuration,
as a consequence of the gauge fixing, $\theta^{'}$ is proportional
to $(\partial_\mu A'_\mu)^2$, which has the vacuum quantum numbers and
mixes with the identity. The expectation value of
$(\partial_\mu A'_\mu)^2$, therefore, diverges as ${1 \over a^4}$
so that $\theta^{'}$ will stay finite, as $a \rightarrow 0$.

\section{Summary and Addendum}
Every step of the usual gauge fixing procedure is affected by subtleties.
The Gribov copies can be moderately dangerous in a simulation but it is
necessary to check their influence in any calculation. The definition of
the gluon field in terms of the links is not a fixed prescription of the
theory but it can be chosen, for example,
in order to satisfy practical requests.

It is also possible to take advantage from this freedom as it has been done
in ref. \cite{pisa,poster} in order  to implement a procedure to fix a generic
covariant gauge on the lattice. The great advantage of a covariant gauge is that
varying the value of the gauge parameter it is possible to check numerically,
 in the calculation
of gauge dependent Green's functions like for example the gluon propagator,
the gauge dependence of the fitted parameters.

After the end of this workshop a thorough study of the lattice covariant
gauges and their applications has been completed \cite{prd}.
 
\section{Acknowledgements}
I thank Valya Mitrjushkin and all the organizers of the workshop
``Lattice Fermions and Structure of the Vacuum'' for the warm hospitality
and the excellent organization.

\end{document}